# Unveiling Deception: Establishing a Taxonomic Framework for Disinformation within Scientific Discourse

Leslie D. McIntosh[1], William White[2], Cynthia Hudson Vitale[1,3]

[1]Digital Science; [2] University of Wisconsin, Milwaukee, [3] Johns Hopkins University

*Corresponding author:
Leslie D. McIntosh
**Email:** leslie@digital-science.com

**Author Contributions:** LM, WW, and CHV designed research and performed research; LM and WW compiled and analyzed data; and LM , WW, and CHV wrote the article
**Competing Interest Statement:** Leslie McIntosh is a full-time employee and Vice President of Research Integrity at Digital Science. Cynthia Hudson Vitale receives a part-time salary from Digital Science. Both provide advice on products related to research integrity.

**This file includes:**
Main Text
Tables 1 to 2

# Abstract

Disinformation spreads among the public and in scientific discourse through the actions of individuals, organizations, and governments that distort scholarly communications, media narratives, and institutional trust. This taxonomy introduces a structured framework and specialized set of definitions to elucidate the key participants, platforms, and strategies employed in the propagation of disinformation. Enhanced comprehension of the mechanisms and pathways of scientific disinformation equips journalists and policymakers with the tools necessary to more effectively recognize and address these issues. The authors developed this taxonomy of disinformation through a multi-faceted approach, encompassing a literature review, expert review, and case study analysis. The literature review revealed a scarcity of taxonomical models amidst prevalent algorithmic detection studies. Subsequently, an expert review process refined our taxonomy through collaborative analysis of twenty-two cases of identified disinformation, categorized by their methods, motives, and impacts. Finally, we validated and fine-tuned our taxonomy through detailed case studies of twelve diverse disinformation instances, assessing the taxonomy's effectiveness in capturing the essential characteristics of each case and making necessary adjustments to ensure its relevance and accuracy in real-world applications.

# Introduction

The creation and spread of disinformation have become commonplace in our daily lives. Over the past decade, significant research has highlighted the rise of disinformation and its proliferation across various populations. A notable study by Barthel et al. in 2016 found that 64% of adults believe fake news stories significantly contribute to public confusion, and 23% admitted to sharing fabricated political stories—sometimes inadvertently and sometimes deliberately. Furthermore, in 2013, the World Economic Forum identified the rapid dissemination of disinformation via social media as one of society's top ten critical trends. The lack of standard definitions and typologies of disinformation further complicates issues. For this research we have adopted the European Commission (2018) definition in which disinformation is defined as any false, inaccurate, or misleading information that is intentionally designed, presented, and promoted to cause public harm or for profit.‘Misinformation’ bears a deceptively close definition to disinformation, but with unintentional motivation from the actor. For example, if a person passes on information unknown to be false but thought to be true (or they didn’t think at all as passed on the information because they agreed with it) - that would be considered misinformation. Intent must be known to differentiate mis and dis-information, which is tricky as any debate in assigning one or the other dilutes the focus on misleading information.

Despite the exponential growth of disinformation, only a few existing classification models seek to develop a common set of terms and definitions, with most focusing broadly on misinformation or 'fake news' and intent.  For example, Zhou and Zhang (2007) organize misinformation in an ontology based on properties such as type, motivation, and content. Wardle (2007) introduces a typology that measures the intent to deceive, while White (2023) classifies scholarly disinformation by authorial intent, differentiating between parodic, opportunistic, or malicious origins. Kapantai, et. al. (2020) conducted a systematic review across 10 different disinformation taxonomies or typologies of disinformation. Looking across these existing frameworks, the authors distilled three independent dimensions and several controlled values per dimension to categorize the types of disinformation.

These existing models do much to build a shared understanding of types of disinformation in forms of popular communication. Yet, they underscore the lack of comprehensive theoretical models for disinformation within academic literature. Our research investigates the development of a systematic taxonomy for dissecting and understanding the complex nature of disinformation, particularly in scientific communication.

Taxonomies provide a systematic way to categorize information based on observable actions and characteristics, offering a universal language that enhances the objectivity of analysis and aids scholars, researchers, and policymakers. This approach not only helps distinguish between various forms of disinformation—such as fabrication, distortion, or omission—but also enables the identification of specific issues within the broader spectrum of disinformation, fostering more effective countermeasures.

# Research questions

We sought to answer the following questions to develop the taxonomy:

1. What are the identifiers of scientific manipulations?
2. Through what entities do scientific manipulations manifest? What are the platforms/mechanisms through which scientific manipulations manifest?
3. Who are the actors? How do we classify scientific manipulations? (Science is defined in the classical sense.)

# Materials & Methods

We conducted a systematic literature review, expert reviews, and case study analyses to develop a comprehensive taxonomy of disinformation.

## Literature review

To appraise the existing literature regarding taxonomies of disinformation, we conducted a systematic database search to cover diverse scholarly works across multiple disciplines. Using a combination of truncation and phrase searching, we used the following search string: (taxonom* or ontolog* or classif*) and (misinformation or disinformation or "fake news"). We searched for the terms within titles and abstracts of research articles and excluded books. We did not filter results by publication date. Our search yielded the following results:

PubMed: 29,469 results; 341 with Title/Abstract Query
Academic Search Complete: 412
Web of Science: 1239
Google Scholar: 181
Dimensions: 2032; 112 for research articles only Few of the search results across these databases propose taxonomical disinformation models. Instead, most of the search results concern the detection of misinformation and disinformation through algorithmic models, with most of these studies concentrated on social media (Bani-Hani, et al., 2021).

### Expert and Case Study Reviews

The authors initially drafted a list of all potential sources of scholarly disinformation. Over 6 weeks, we iteratively reviewed, classified, and refined the identified sources to eventually form the taxonomy of disinformation. We began by collecting twenty-two examples of content that could be considered disinformation that would form the basis of the case study reviews. We collected these examples based on the following criteria:

1. Original articles or online documents must be publicly available;
2. A statement or verified document was published indicating a lapse in the conduct of responsible research and that statement indicates an instance of disinformation (e.g., identified as false information)

These cases were then categorized based on the methods used, motivations, and effects of spreading such content. Through regular meetings and discussions, we refined the categorical framework. This collaborative and iterative process enabled us to develop a comprehensive taxonomy grounded in the expertise of the author team.

To evaluate the comprehensiveness and utility of our proposed taxonomy of disinformation, we conducted an applied case study analysis using twelve recent examples of disinformation campaigns or content. The cases were selected to encompass various sources, motivations, formats, and topics. We examined the available evidence and historical records for each case to classify the disinformation according to our taxonomy categories. We assessed how well each taxonomy component captured the meaningful attributes of the case and identified any areas requiring refinement or additional categories. By systematically applying the draft taxonomy to these test cases, we aimed to validate the framework's ability to provide a nuanced and accurate representation of real-world disinformation. This case-based approach enabled us to refine the taxonomy.

## Results

Our work shows that to gain a more comprehensive understanding of disinformation in scientific scholarship, it is essential to address three fundamental questions (in any order): Who or what are the perpetrators of these acts (Actors)? From what sources does the disinformation emanate (Outlets)? And by what means is the disinformation propagated (Methods)? (Table 1)

Actors are the individuals, organizations, and governments actively participating in the spread of disinformation. However, these actors can only spread disinformation through outlets such as journals, events, media, and institutions to carry their messages. Lastly, methods used to spread disinformation include deceiving scholarly communication, gaming mainstream media, and leveraging the judicial system. Table 2 (A-C) represents the details of these parent categories.

<<Insert tables 1 & 2>>

## Findings: Case Studies

The two authors (LMc and WW) classified the twelve case studies and had a general agreement on *who* perpetrated the disinformation and *where* it flowed. Classifying the methods of *how* disinformation flowed proved a bit trickier. The two authors agreed on six cases, had near agreement on four, and did not match classifications on two. Part of this challenge was that multiple reasons could be assigned within each category. After discussions, LMc and WW agreed on classifications for all case studies. The table below lists the case studies reviewed for this analysis.

| | **Case Study** |
|---|---|
| 12 | Retracted COVID-19 Paper Misrepresents Data |
| 3 | Expression of Concern for Wrong Test |
| 7 | Falsified Data and Stanford President Resignation |
| 17 | Donor pressure in science |
| 1 | Conference Panel Cancellation |
| 4 | Undisclosed Affiliations - Dr. Charles Lieber |
| 11 | Withdrawn COVID-19 Preprint on Pharmacological Outcomes |
| 15 | Honesty Researcher Fabricates Data |
| 21 | Google AI Study in Nature Flagged |
| 13 | Retracted COVID-19 Paper |
| 19 | Fraudulent Studies from Medical Professional Joachim Boldt |
| 20 | Publishing Satire - Sokal Affair |

### Example Case Studies

In one case (ID 12), Retracted COVID-19 Paper Misrepresents Data, scholars Walach, Klement, and Aukema used a peer-reviewed journal to misrepresent science - specifically the data.

Even though the article was retracted, Dr. Walach placed the information on his website, bypassing the scholarly investigation to publicly present his work - potentially misleading readers to what may be seen as upstanding research. In this case study, the individual author (who) was responsible for misrepresenting their sources (how) disinformation. They chose to spread it through scholarly publications (where).

In Donor pressure in science (ID 17), also known as the SOKAL III ‘hoax, a journal article was retracted due to fabricated data and a false persona. The anonymous actors presented faked research as upstanding scholarship to ‘prove’ the junk science could get through the peer-review cycle.

This case exemplifies how the taxonomy can be used to identify gaps in knowledge more quickly. Because the instigator(s) remain(s) anonymous, we cannot answer who perpetrated this disinformation. Because they also state other papers exist but those papers have not been identified, then the intention of the instigators also remains blurred. They could be scholars who wanted to make the research arena more rigorous by calling attention to a weak part in the scholarly checks, but they should have not allowed any paper to be published. They could be scholars or non-scholars intent on weakening the scientific ecosystem by further corroding trust in scientific checks. hey could have also been working with an organization to spread disinformation. Nonetheless, they used deceptive communication tactics to misrepresent their persona and science through a publication in a scholarly journal. In this case study, (who) is unverified, yet responsible for creating a false persona (how) and false data (how) and publishing it in a scholarly journal (where).

Undisclosed Affiliations - Dr. Charles Lieber (ID 4) presented an interesting test of the taxonomy and clarified its purpose. In this case, Dr. Lieber, a Harvard chemist did not disclose financial and research ties to another university in another country. The taxonomy is designed to classify scholarly disinformation not detecting something like espionage. However, after this case was inspected and understood, we modified the taxonomy to include this possibility. The taxonomy now accounts for institutions like academia, corporations, and government as locations where disinformation could be distributed - not only where it originated from. In this case was a scholar (who), misrepresenting himself by concealing affiliations (how) and used his academic institution in collaboration with another government (where) to train researchers. This case also highlights that disinformation manifests through a person and their affiliations, not their scholarship.

## Limitations

Despite their numerous advantages, taxonomies also come with certain limitations. One major challenge is the potential for subjectivity in categorization. Researchers may interpret and classify the same information differently depending on perspectives and biases. Taxonomies can clarify this information by having clear and objective criteria for classification, but regular validation exercises should be conducted to ensure consistency.

Another limitation is the dynamic nature of disinformation. Taxonomies, while valuable, can quickly become outdated as new tactics and technologies for spreading false information emerge. Maintaining a taxonomy's relevance requires constant monitoring and adaptation to keep pace with the evolving landscape of disinformation.

The taxonomy does not specifically address the disinformation that may result from using artificial intelligence (AI). Conducting further research into the various types of disinformation that could emanate from AI and other emerging technologies is a crucial next step to mitigate these risks.

Additionally, taxonomies may oversimplify complex phenomena. Disinformation often operates in shades of grey, with hybrid forms and subtle variations that can be challenging to categorize neatly. Researchers must be cautious not to force complex information into rigid taxonomy structures, as this could lead to a loss of nuance and a failure to capture the intricacies of disinformation campaigns.

## Discussion

This research systematically explores the mechanisms of disinformation in scientific scholarship by addressing three critical questions: the actors involved, the outlets for disinformation, and the methods by which it is spread. Through an analysis of twelve diverse case studies, our findings illuminate the complexities and nuances of identifying and classifying disinformation within the publishing realm.

The practical application of a taxonomy further extends to journal editors and researchers. For instance, during the COVID-19 pandemic, a taxonomy-based checklist could have expedited the review of manuscripts and preprints, potentially reducing the publication of disinformation in scholarly literature. A taxonomy is especially important as issues around scientific and research integrity continue to significantly concern the research enterprise (White House, 2023). By clarifying the types, sources, and goals of varying disinformation, the taxonomy can inform legislative and regulatory efforts to limit the spread and impact of disinformation. Further, by naming common issues and tactics, a taxonomy enhances monitoring and analysis by enabling scholars to tag, track, and study disinformation campaigns and trends over time. It has broad implications outside the scholarly sphere. Information literacy programs, essential for verifying sources and mitigating misinformation, can benefit from this taxonomy. Educators can use it to develop curricula that improve students' skills in identifying and critically evaluating disinformation, enhancing their online research and information consumption practices.

### Actors and Their Roles

The actors identified in our case studies—from individual scholars to organizations and even governments—highlight the varied sources of disinformation. The diversity among actors suggests that the impacts of disinformation are varied and complex. Clarifying the actors allows discussion to move more quickly to their impacts.

For instance, in the case of the retracted COVID-19 paper on vaccinations (Case ID 12), the disinformation originated from individual authors and a peer-reviewed publication on a scholarly platform onto a non-scholarly website detached from the scholarly record. Thus, individual actors perpetuate misinformation who can have significant impacts when they misuse scientific communication platforms.

Donor pressure in science (Case ID 17), further illustrates the complex nature of disinformation, where actors may not be known and could be individuals, organizations, and/or governments working independently or colluding together. This dual-layered disinformation - using fabricated data and false personas - not only distorts scientific facts but also manipulates the identity and credibility of the purported scientists, further complicating the detection processes. However, the taxonomy allows us to more quickly classify actors and identify gaps (i.e., knowing the perpetrators).

### Outlets and Methods as Points for Disinformation

Our findings underscore the critical role of outlets in disinformation dissemination. Journals, conferences, and online platforms are among the primary avenues for propagating erroneous information. The modification of our taxonomy to include institutions such as academia, corporations, and government entities as outlets (as observed in the case of Dr. Charles Lieber, Case ID 4) reflects the evolving understanding of how deeply intertwined these entities can be in the spread of disinformation.

The methods of disinformation, including deceptive scholarly communication and the manipulation of media and legal frameworks, reveal the sophisticated strategies employed to influence public and academic discourse. The complexity in classifying these methods, as evidenced by the variability in agreement between our authors, indicates the challenges inherent in dissecting the nuances of disinformation tactics.

### Impacts

Yet, the complexities of classifying disinformation should not deter us as the consequences for spreading disinformation are too great. Perpetuating opinion under the guise of upstanding scholarship harms human health. As seen in the case of Wallach, et al (Case 12), where they falsely claimed a link between vaccinations and poor health outcomes, the disinformation crowds and muddles the conversations and decisions around health. COVID was a real pandemic jeopardizing lives and vaccinations offer a potential solution, with risks. In this recent study the impact has been quantified: “Allen et al. suggest that exposure to a single piece of vaccine misinformation reduced vaccination intentions by ~1.5 percentage points.” (van der Linden & Kyrychenko, 2024).

The impacts of ID 4 involving Charles Lieber are intriguing. Lieber was convicted and served time for lying to US federal authorities about his involvement with the Chinese Thousand Talent Program. Initially presented as a Platform for academic exchanges, this program was later revealed to be a deceptive initiative by the Chinese government to steal state intellectual property from one nation and transfer it to another.

The most significant challenge of all disinformation tactics is their potential to undermine science. Whether intentional or not, as both Wallach and SOKAL III have shown, disinformation tactics have eroded the public's trust in science, corroding the perception of science as a search for truth and understanding. This erosion, akin to other 'disruptions,' paves the way for the manipulation of fiction to sound like truth.

## Conclusion and Recommendations

These findings have profound implications for developing policies and practices to mitigate scientific disinformation. Enhancing the rigor of peer review, increasing transparency in author affiliations and funding disclosures, and fostering a culture of integrity within academia are essential steps toward safeguarding the scientific enterprise.

With the development of this taxonomy, the scientific communications enterprise can more easily classify and collect data on disinformation. In addition to encouraging the further development of this taxonomy, we suggest the following recommendations for various stakeholders in the scholarly communications environment to mitigate disinformation.

- Enhanced Peer Review Processes: Publishers should implement robust peer review mechanisms that include checks for data veracity and reproducibility. Consider involving a wider range of experts and employing open peer review to increase transparency. For special issues, assign an independent co-editor to mitigate bias.
- Education and Training: Institutions, publishers, and scholarly societies should develop and provide training resources for researchers, reviewers, and editors on identifying and combating disinformation. These activities could include workshops, webinars, and online courses on critical thinking, data verification, and digital literacy.
- Collaboration with Fact-checking Organizations: Publishers and journalists should establish partnerships with professional fact-checking organizations to validate research findings that gain significant public attention or could have major public policy implications.
- Promotion of Open Access and Data Sharing: All stakeholders should encourage open access to research articles and data. This would enable more comprehensive scrutiny and verification by the global research community, which can help quickly identify and correct misleading or manipulated information.
- Strengthen Editorial Standards: Publishers should update and enforce stringent editorial policies that require thorough citation of sources, declaration of all conflicts of interest, and transparency about research funding.
- Public Engagement and Communication: Scholars and institutions should engage with the public through outreach and education initiatives to improve understanding of how scholarly research is conducted, peer-reviewed, and published, enhancing public trust and resilience against disinformation.
- Rapid Response Mechanisms: Publishers and journalists should create protocols for quickly addressing instances where published research is found to be compromised or used to spread disinformation, including timely corrections, retractions, and clear communications with the public.
- Monitoring and Auditing: Publishers and journalists should regularly monitor and audit published content to ensure ongoing compliance with standards. Use analytics to detect patterns that might suggest problematic research practices or targeted disinformation campaigns.
- International Cooperation: Researchers and scholars should work across borders with international bodies to develop standardized approaches and share best practices for combating disinformation in scholarly communications.

By continuing to refine our taxonomical framework and applying it across a broader spectrum of cases, we can better equip the academic community to identify and counteract the pernicious effects of disinformation, thereby preserving the integrity of scientific discourse.

**Tables**

**Table 1:** Scholarly Disinformation Taxonomy.

| Scholarly Disinformation Taxonomy | | |
|---|---|---|
| Category | Subcategory | Definition and examples |
| Actors (Who) | Individuals | The individual who published or created the disinformation. Includes 2 types, for example: a scholar/researcher or non-scholar. |
| | Organizations | The organization that published or created the disinformation. Includes 7 types, for example: a lobby organization or a mimic organization. |
| | Governments | A government supported or created the disinformation. Includes 3 types. |
| Outlets (Where) | Journals | Indicates if a journal was the mechanism for distributing the disinformation. Includes 3 types, for examples: predatory journals. |
| | Events | Indicates if an event was the mechanism for distribution the disinformation. Includes 2 types, for example: scientific conferences. |
| | Media | Indicates if the media was the mechanism for distribution of the mis or disinformation. Include 4 types, for example: social media or websites. |
| | Institutions | Indicates if an institution was the mechanism for distribution of the disinformation. Includes 3 types, for example: a political action committee. |
| Methods (How) | Deceiving Scholarly Communication | Deceiving scholarly communication was how the disinformation took place. Includes 3 events and 17 sub-events, for example: misrepresenting science/scholarship by fabricating data. |
| | Gaming Mainstream Media | Gaming mainstream media was how the disinformation materialized. Includes 6 events, for example: emotionally charged headlines and spreading conspiracy theories. |

| | |
|---|---|
| Leveraging Judicial System | Leveraging the judicial system was how the disinformation was perpetuated. Includes 4 events, for example: misusing the law and lobbying activities. |

**Table 2. Categories with subcategories and types of disinformation taxonomy (A) Actors, (B) Outlets, and (C) Methods.**

| (A) Actors (Who) | |
|---|---|
| Subcategory | Types |
| Individuals | Scholar/researcher<br>Non-scholar |
| Organizations | Academic institutions<br>Think Tanks<br>Lobbying Organizations<br>Corporations<br>Mimim Organization<br>Extremist groups<br>News Outlets |
| Governments | State media<br>Politicians<br>Military Officials |

| (B) Outlets (Where) | |
|---|---|
| Subcategory | Events |
| Journals | Established<br>Vanity/Mock<br>Predatory |
| Events | Scientific Conference<br>Misleading (Predatory, Stacked) |
| Media | Social<br>News outlet<br>'Entertainment' media<br>Websites |
| Institutions | Academic<br>Corporate<br>Government (Judicial) |

| (C) Methods (How) | | |
|---|---|---|
| Subcategory | Events | Sub-events |
| Deceiving Scholarly Communication | Misrepresenting Person | Claiming false credentials<br>False persona<br>Concealing affiliations and/or political ties |
| | Misrepresenting Science/Scholarship | Misrepresenting sources<br>Cherry-picking evidence<br>Falsification<br>Fabricating data/information<br>Plagiarism<br>Non-reproducible results |
| | Manipulation Publishing Process | Becoming guest editors<br>Creating vanity journals |
| | | Manipulating peer review<br>Relying on predatory or mock journals/publishers<br>Manipulate/inflate scientific metrics<br>Preying on weak publishing process |
| Gaming Mainstream Media | Spreading conspiracy theories<br>Manufacturing controversy<br>Emotionally charged headlines<br>Distorted Images/Visuals<br>Hate Speech<br>Amplify fringe voices | |
| Leveraging Judicial System | Lawsuits<br>Misusing Law<br>Lobbying Activities<br>Government withholding funding | |